# Orbital-Angular-Momentum Versus MIMO: Orthogonality, Degree of Freedom, and Capacity

Haiyue Jing[†], Wenchi Cheng[†], Xiang-Gen Xia[‡], and Hailin Zhang[†]
[†]State Key Laboratory of Integrated Services Networks, Xidian University, Xi'an, China
[‡]Department of Electrical and Computer Engineering, University of Delaware, Newark, DE 19716
E-mail: {[†]hyjing@stu.xidian.edu.cn, [†]wccheng@xidian.edu.cn, [‡]xxia@ee.udel.edu, [†]hlzhang@xidian.edu.cn}

*Abstract*—The plane wave based wireless communications have becoming more and more matured, along with the well utilization of the traditional resources such as time and frequency. To further increase the capacity for rapidly increasing capacity demand of wireless communications, it is potential to use the twist wave, which has the orbital angular momentum (OAM). In this paper, we discuss the OAM based wireless communications in the aspect of orthogonality, degree of freedom (DoF), and capacity, where both the transmitter and the receiver use uniform circular array (UCA) antennas. In particular, we compare OAM based wireless communications with multiple-input-multiple-output (MIMO) based wireless communications in terms of DoF and capacity. Numerical results are presented to validate and evaluate that the DoF of OAM based wireless communications is greater than or equal to that of correlated MIMO based wireless communications when the transmitter and the receiver antennas are aligned well. The OAM based wireless communications can achieve larger capacity than the correlated MIMO in high signal-to-noise ratio (SNR) region under line-of-sight scenario.

## I. Introduction

During the past few decades, different dimensional resources have been exploited to meet the increasing capacity demand in wireless communications [1]. However, the resources, such as time and frequency, have been extensively explored. It is now very difficult to increase capacity using the corresponding techniques such as time-division-multiple-access and frequency-division-multiple-access.

In fact, the traditional wireless communications only use the linear momentum while the orbital angular momentum (OAM) is not attached importance to or well used [2]–[4]. Recently, academic researches have paid much attention to the OAM, which is a result of a signal possessing helical phase front. OAM contains a great number of topological charges, i.e., OAM-modes. Beams with different OAM-modes are orthogonal with each other and they can be multiplexed/demultiplexed together, thus increasing the capacity without relying on the traditional resources such as time and frequency.

So far, many experiments have been conducted to validate that OAM can be used for data transmission. The authors of [5] conducted an outdoor experiment, where two OAM-modes (0 and 1) were multiplexed and demultiplexed using the same frequency band, to show the feasibility of OAM based wireless communications. The authors of [6] developed a high-capacity millimeter wave communication system with OAM multiplexing. The authors of [7] experimentally demonstrated that OAM can be used to increase capacity for wireless communications within Rayleigh distance. Also, OAM based wireless communications have received much attention in the aspect of mode detection [8], mode decomposition [4], axis estimation and alignment [9], mode modulation [10], OAM-beams converging [11], and mode hopping [12], etc.

For high capacity transmission in wireless communications, uniform circular array (UCA) is considered as one promising antenna architecture for OAM based wireless communication system due to its flexibility in radiating multiple OAM beams with different OAM-modes simultaneously [13]. The authors of [14] showed that UCA is better than the radial array and the tangential array to generate the desired OAM carrying beam with linear excitation. However, because of the same antenna structure, the UCA based radio vortex transmission is usually compared with the multiple-input-multiple-output (MIMO) based wireless communications. It is not very clear that which one can obtain the largest capacity between OAM based wireless communications and MIMO based wireless communications under the UCA antenna structure.

This work was supported in part by the National Natural Science Foundation of China (No. 61401330), the 111 Project of China (B08038), the Young Elite Scientists Sponsorship Program By CAST, and the Fundamental Research Funds for the Central Universities.

In this paper, we analyze the orthogonality, degree of freedom (DoF), and capacity for OAM based wireless communications, where both the transmitter and receiver are UCA antennas and they are parallel and coaxial. Because of the small size of UCA, the antenna-elements usually form the correlated MIMO. To compare the DoF and capacity of the OAM based wireless communications with those of the correlated MIMO based wireless communications, we first calculate the spatial correlation matrix corresponding to the receiver of MIMO based wireless communications. Then, we derive the DoF and capacity based on the spatial correlation matrix. We conduct extensive simulations to validate and evaluate that the OAM based wireless communications is better than the correlated MIMO based wireless communications in terms of DoF and capacity under high signal-to-noise ratio (SNR) region.

The rest of this paper is organized as follows. Section II gives the OAM and MIMO based wireless communications models. Section III derives the channel models for OAM based wireless communications and MIMO based wireless communications, respectively. Section IV analyzes the orthogonality of OAM based wireless communications. We also compare the DoF and capacity of OAM based wireless communications with those of the MIMO based wireless communications. Section V evaluates the capacities and the ratio of DoF related to the OAM based wireless communications to the DoF related to MIMO based wireless communications. The paper concludes with Section VI.

## II. SYSTEM MODEL FOR UCA BASED WIRELESS COMMUNICATIONS

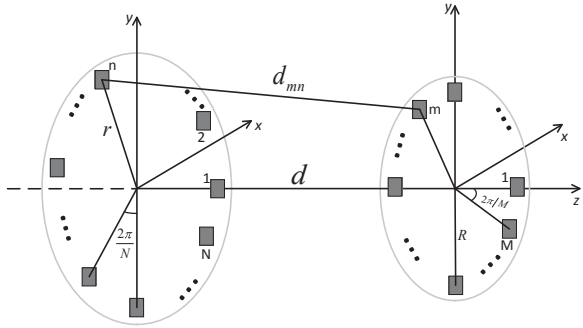

Fig. 1. The system model for the OAM or correlated MIMO based wireless communications.

Figure 1 shows the system model for OAM based wireless communications or correlated MIMO based wireless communications, where the transmitter and receiver are aligned with each other and the antenna-elements of transmitter and receiver are uniformly around the perimeter of the circle. We denote by $r$ and $R$ the radii of the transmitter and receiver, respectively, where $r$ and $R$ can be different. The notation $d$ represents the distance between the center of transmitter and the center of receiver while $d_{mn}$ represents the distance from the $n$th antenna-element at the transmit UCA to the $m$th antenna-element at the receive UCA. There are $N$ and $M$ antenna-elements at the transmit UCA and receive UCA, respectively. If the system corresponds to the OAM based wireless communications, the antenna-elements are fed with the same input signal but with a successive delay from antenna-element to antenna-element such that after a full turn the phase has been incremented by an integer multiple $l$ of $2\pi$, where $l$ represents the order of OAM-mode. If the system refers to the MIMO based wireless communications, it is not necessary for the antenna-elements to be fed the same signals and the signals are usually different. In the following, we study the orthogonality, DoF, and capacity of OAM based wireless communications. We also compare the DoF and capacity of OAM based wireless communications with those of MIMO based wireless communications.

## III. CHANNEL MODEL FOR UCA ANTENNA STRUCTURE BASED WIRELESS COMMUNICATIONS

We denote by $h_{mn}$ the channel gain from the $n$th antenna-element on the transmit UCA to the $m$th antenna-element on the receive UCA. Then, $h_{mn}$ can be written as follows:

$$h_{mn} = \frac{\beta \lambda e^{-j\frac{2\pi}{\lambda}d_{mn}}}{4\pi d_{mn}}, \tag{1}$$

where $\beta$ denotes all relevant constants such as attenuation and phase rotation caused by antennas and their patterns on both sides. We denote by $c$ the distance between the projection of the $n$th antenna-element on the transmit UCA in the receive plane and the $m$th antenna-element on the receive UCA. Then, we have

$$c = \sqrt{r^2 + R^2 - 2rR\cos(\phi_n - \psi_m)}, \tag{2}$$

where $\phi_n = 2\pi(n-1)/N$ is the azimuthal angular and also the angular position on a plane perpendicular to the axis of propagation, at the transmit UCA corresponding to the $n$th antenna-element. The parameter $\psi_m = 2\pi(m-1)/M$ is the azimuthal angle for receive UCA corresponding to the $m$th antenna-element. Then, we can derive $d_{mn}$ as follows:

$$\begin{aligned} d_{mn} &= \sqrt{d^2 + c^2} \\ &= \sqrt{d^2 + r^2 + R^2 - 2rR\cos(\phi_n - \psi_m)}. \end{aligned} \tag{3}$$

Because of $d_{mn} \gg r$ and $d_{mn} \gg R$, we can make approximation for $d_{mn}$ at the denominator and numerator in Eq. (1). For the denominator, we use $d_{mn} \approx d$. For the numerator, $d_{mn}$ can be rewritten

according to $\sqrt{1-x} \approx 1-x/2$ as follows:

$$
\begin{aligned}
d_{mn} &= \sqrt{d^2+r^2+R^2}\sqrt{1-\frac{2rR\cos(\phi_n-\psi_m)}{d^2+r^2+R^2}} \\
&\approx \sqrt{d^2+r^2+R^2}\left[1-\frac{rR\cos(\phi_n-\psi_m)}{d^2+r^2+R^2}\right] \\
&= \sqrt{d^2+r^2+R^2}-\frac{rR\cos(\phi_n-\psi_m)}{\sqrt{d^2+r^2+R^2}}. \quad (4)
\end{aligned}
$$

A. *Channel Model for MIMO Based Wireless Communications*

Taking Eq. (4) into Eq. (1), we can rewrite $h_{mn}$ as follows:

$$
\begin{aligned}
h_{mn} = &\frac{\beta\lambda e^{-j\frac{2\pi}{\lambda}\sqrt{d^2+r^2+R^2}}}{4\pi d} \\
&\times \exp\left[\frac{j2\pi rR}{\lambda\sqrt{d^2+r^2+R^2}}\cos(\phi_n-\psi_m)\right]. \quad (5)
\end{aligned}
$$

Because the size of UCA is relatively small, the antenna-elements usually form correlated MIMO. We assume the signal follows the Gaussian angle distribution and the angular distribution function can be re-expressed as follows [15]:

$$
p(\theta) = \frac{\kappa}{\sqrt{2\pi}\delta}e^{-\frac{(\theta-\vartheta)^2}{2\delta^2}}, \quad -\pi+\vartheta \leq \theta \leq \pi+\vartheta, \quad (6)
$$

where $\vartheta$ and $\delta$ are the mean direction of arrival and the standard deviation of the distribution, respectively. The parameter $\kappa$ is the normalization factor given as follows:

$$
\kappa = \frac{1}{\frac{2}{\sqrt{\pi}}\int_0^{\frac{2}{\sqrt{\pi}\delta}}e^{-t^2}dt}. \quad (7)
$$

When $\delta$ is small, $\kappa$ is very close to 1.

Then, the mathematical expression for spatial correlation function, denoted by $\rho_{u,v}$, between the $u$th antenna-element and the $v$th antenna-element at the receive UCA is given as follows:

$$
\rho_{u,v} \approx \kappa\exp\left[-j\xi\sin(\nu+\vartheta)-\frac{[\xi\delta\cos(\nu+\vartheta)]^2}{2}\right], \quad (8)
$$

where $1 \leq u \leq M$ and $1 \leq v \leq M$. The notations $\xi$ and $\nu$ are given by

$$
\xi = \frac{2\pi R}{\lambda}\sqrt{2-2\cos(\psi_u-\psi_v)} \quad (9)
$$

and

$$
\begin{cases} \sin\nu = \frac{\cos\psi_u-\cos\psi_v}{\sqrt{2-2\cos(\psi_u-\psi_v)}}; \\ \cos\nu = \frac{\sin\psi_u-\sin\psi_v}{\sqrt{2-2\cos(\psi_u-\psi_v)}}, \end{cases} \quad (10)
$$

respectively.

The channel matrix can be expressed as follows [16]:

$$
\boldsymbol{H}_{\text{MIMO}} = \boldsymbol{G}_r^{1/2}\boldsymbol{H}_g\boldsymbol{G}_t^{1/2}, \quad (11)
$$

where $\boldsymbol{H}_g$ is a matrix with i.i.d Gaussian entries. $\boldsymbol{G}_t$ and $\boldsymbol{G}_r$ are the spatial correlation matrices at the transmit UCA and receive UCA, respectively. According to Eq. (8), we have the spatial correlation matrix at the receiver as follows:

$$
\boldsymbol{G}_r = \begin{bmatrix} \rho_{0,0} & \rho_{0,1} & \cdots & \rho_{0,M-1} \\ \rho_{1,0} & \rho_{1,1} & \cdots & \rho_{1,M-1} \\ \vdots & \vdots & \ddots & \vdots \\ \rho_{M-1,0} & \rho_{M-1,1} & \cdots & \rho_{M-1,M-1} \end{bmatrix}. \quad (12)
$$

Through the process, which is similar to the process of getting the spatial correlation matrix $\boldsymbol{G}_r$ at the receive UCA, we can obtain the spatial correlation matrix $\boldsymbol{G}_t$ at the transmit UCA.

B. *Channel Model for OAM Based Wireless Communications*

Before spatially sampling at the receiver, the signal received at the $m$th antenna-element of the receive UCA, denoted by $r_m$, can be derived as follows:

$$
\begin{aligned}
r_m &= \sum_{l=\frac{1-N}{2}}^{N/2}\sum_{n=1}^{N}h_{mn}\frac{1}{\sqrt{N}}s_l e^{j\frac{2\pi(n-1)}{N}l} \\
&= \sum_{l=\frac{1-N}{2}}^{N/2}\widetilde{h}_{ml}s_l, \quad (13)
\end{aligned}
$$

where $s_l$ is the signal corresponding to the $l$th OAM-mode. The parameter $\widetilde{h}_{ml}$, which is the channel gain for the $l$th OAM-mode corresponding to the transmit UCA and the $m$th antenna-element at the receive UCA, can be given as follows:

$$
\begin{aligned}
\widetilde{h}_{ml} = &\sum_{n=1}^{N}\frac{\beta\lambda e^{-j\frac{2\pi}{\lambda}\sqrt{d^2+r^2+R^2}}}{4\pi d\sqrt{N}}e^{j\frac{2\pi(n-1)}{N}l} \\
&\times \exp\left[\frac{j2\pi rR}{\lambda\sqrt{d^2+r^2+R^2}}\cos(\phi_n-\psi_m)\right]. \quad (14)
\end{aligned}
$$

When $N$ is relatively large, we have Eq. (15), where $\phi_n = 2\pi(n-1)/N$ and $\widetilde{\phi}_n$ is a continuous variable of $\phi_n$ ranging from 0 to $2\pi$. Taking Eq. (15) into Eq. (14), we can approximate $\widetilde{h}_{ml}$ as follows:

$$
\begin{aligned}
\widetilde{h}_{ml} \approx &\frac{\beta\lambda\sqrt{N}e^{-j\frac{2\pi}{\lambda}\sqrt{d^2+r^2+R^2}}e^{j\frac{2\pi(m-1)}{M}l}}{4\pi d(-j)^l} \\
&\times J_l\left(\frac{2\pi rR}{\lambda\sqrt{d^2+r^2+R^2}}\right), \quad (16)
\end{aligned}
$$

where

$$
J_l(\alpha) = \frac{(-j)^l}{2\pi}\int_0^{2\pi}e^{jl\tau}e^{j\alpha\cos\tau}d\tau \quad (17)
$$

is the $l$-order Bessel function. Eq. (16) shows that for different antenna-elements on the receive UCA, the corresponding channel gains $\widetilde{h}_{ml}$ follow the Bessel

$$\frac{(-j)^l}{N} \sum_{n=1}^{N} \exp\left[j\left(\phi_n - \psi_m\right)l\right] \exp\left[\frac{j2\pi rR}{\lambda\sqrt{d^2 + r^2 + R^2}} \cos\left(\phi_n - \psi_m\right)\right]$$

$$\approx \frac{(-j)^l}{2\pi} \int_0^{2\pi} \exp\left[j\left(\widetilde{\phi}_n - \psi_m\right)l\right] \exp\left[\frac{j2\pi rR}{\lambda\sqrt{d^2 + r^2 + R^2}} \cos\left(\widetilde{\phi}_n - \psi_m\right)\right] d\widetilde{\phi}_n$$

$$= \frac{(-j)^l}{2\pi} \int_{-\psi_m}^{2\pi-\psi_m} \exp\left[j\left(\widetilde{\phi}_n - \psi_m\right)l\right] \exp\left[\frac{j2\pi rR}{\lambda\sqrt{d^2 + r^2 + R^2}} \cos\left(\widetilde{\phi}_n - \psi_m\right)\right] d(\widetilde{\phi}_n - \psi_m)$$

$$= J_l\left(\frac{2\pi rR}{\lambda\sqrt{d^2 + r^2 + R^2}}\right). \tag{15}$$

---

distribution with different orders. The azimuthal angle is continuous when the radio vortex signal is transmitted by the transmit UCA. At the receiver, the item $\exp\{j\varphi l\}$ turns to $\exp\{j[2\pi(m-1)/M]l\}$ after spatially sampling. Then, we derive the channel gain, denoted by $\widetilde{h}_l$, corresponding to the $l$th OAM-mode from the transmit UCA to the receive UCA as follows:

$$\widetilde{h}_l = \frac{\beta\lambda\sqrt{N}e^{-j\frac{2\pi}{\lambda}\sqrt{d^2+r^2+R^2}}e^{j\varphi l}}{4\pi d(-j)^l} \times J_l\left(\frac{2\pi rR}{\lambda\sqrt{d^2 + r^2 + R^2}}\right). \tag{18}$$

Observing Eq. (18), we can find that the discrete azimuthal angle in terms of $n$ turns to be the continuous azimuthal angle in terms of $\varphi$ along with the OAM signal transmitting (spatial DA process) and the transmit UCA makes the OAM signal like going through the Bessel-form channel.

We denote by $h_l$ the channel gain without the item $e^{j\varphi l}$ as follows:

$$h_l = \frac{\beta\lambda\sqrt{N}e^{-j\frac{2\pi}{\lambda}\sqrt{d^2+r^2+R^2}}}{4\pi d(-j)^l} J_l\left(\frac{2\pi rR}{\lambda\sqrt{d^2 + r^2 + R^2}}\right). \tag{19}$$

Clearly, $h_l$ is independent of $n$ and $m$. The channel matrix can be written as follows:

$$\boldsymbol{H}_{\text{OAM}} = \text{diag}\left[h_{\frac{1-N}{2}}, \cdots, h_0, \cdots, h_{\frac{N}{2}}\right]. \tag{20}$$

## IV. PERFORMANCE OF OAM AND MIMO BASED WIRELESS COMMUNICATIONS

### A. Orthogonality of OAM-Modes

An EM wave with a helical phase front $\exp(i\varphi l)$ carries an OAM-mode related to $l$, where $\varphi$ is the azimuthal angle and $l$ is an unbounded number (referred as the order of OAM-mode). The phase front of an OAM-mode twists along the propagation direction. When the orders of OAM-modes are integers, the OAM-modes are orthogonal with each other because we have

$$\int_0^{2\pi} s_{q_1} e^{j\varphi q_1} s_{q_2}^* e^{-j\varphi q_2} d\varphi = \begin{cases} 2\pi|s_{q_1}|^2, & q_1 = q_2; \\ 0, & q_1 \neq q_2, \end{cases} \tag{21}$$

where $q_1$ and $q_2$ are indexes of OAM-modes. The parameters $s_{q_1}$ and $s_{q_2}$ refer to complex signals. The notation $(\cdot)^*$ represents the conjugate operation.

When the order of OAM-mode is a non-integer, we can decompose the non-integral OAM-mode into the sum of Fourier series as follows:

$$\exp(j\varphi l) = \frac{\exp(j\pi l)\sin(\pi l)}{\pi} \sum_{q=-\infty}^{+\infty} \frac{\exp(j\varphi q)}{l-q}. \tag{22}$$

However, it is very difficult to demultiplex OAM beams with different OAM-modes if the orders of OAM-modes are not integers, we focus on the OAM-modes when the orders are integers.

For MIMO based wireless communications, there exists interference among the signals, i.e., the signals are not orthogonal with each other.

### B. DoF of OAM and MIMO Based Wireless Communications

With a limited number of antenna-elements equipped equidistantly around the perimeter of an UCA, it can generate multiple waves with different OAM-modes, which are orthogonal with each other.

We assume that there are $N$ and $M$ antenna-elements equipped at the transmit UCA and receive UCA, respectively. The azimuthal angle of the $n$th antenna-element can be calculated by $\exp[j2\pi(n-1)l/N]$ and we have $[(1-N)/2]+1 \leq l \leq [N/2]$, where $[\cdot]$ represents the rounding operation.

*Lemma 1:* The maximum number of OAM-modes for OAM based wireless communications is equal to $\min\{N, M\}$.

*Proof:* When $[(1-N)/2]+1 \leq l \leq [N/2]$, we can easily obtain that there is no interference for different OAM-modes at the transmitter. When $l > [N/2]$ or $l < [(1-N)/2]+1$, the transmit UCA emits the OAM waves with interference. Because there is no difference to our discussion for the parity of the number of antenna-elements $N$ and $M$, we consider $N$ and $M$ both are even numbers in the following.

If $l > N/2$, we can rewrite $l$ as follows:

$$l = \frac{aN}{2} + b, \tag{23}$$

$$C_{\text{OAM}} = \sum_{l=\frac{1-N}{2}}^{N/2} B \log_2 \left(1 + \frac{P|h_l|^2}{\sigma_l^2 N}\right) = \sum_{l=\frac{1-N}{2}}^{N/2} B \log_2 \left[1 + \frac{P\beta^2\lambda^2}{16\pi^2 d^2 \sigma_l^2} J_l^2 \left(\frac{2\pi rR}{\lambda\sqrt{d^2+r^2+R^2}}\right)\right]. \quad (29)$$

where $a$ ($a \geq 1$) and $b$ ($0 \leq b < N/2$) are integers. Then, the azimuthal phase corresponding to the $n$th antenna-element at the transmit UCA can be rewritten as follows:

$$\exp\left[j\frac{2\pi(n-1)}{N}l\right] = \exp\left[j\frac{2\pi(n-1)}{N}\left(\frac{aN}{2}+b\right)\right]$$
$$= \begin{cases} \exp\left[j\frac{2\pi(n-1)}{N}b\right], & \text{if } a \text{ is an odd number;} \\ \exp\left[j\frac{2\pi(n-1)}{N}(b-\frac{N}{2})\right], & \text{if } a \text{ is an even number.} \end{cases} \quad (24)$$

Observing Eq. (24), we can obtain that there exist interfering OAM waves for the transmit UCA. The number of interfering OAM waves increases as $l$ increases. Thus, it is very difficult for the receiver to decompose the signals with different OAM-modes from the transmitter.

On the other hand, different OAM-modes can be demultiplexed by integrating the complex field weighted with a phase term $e^{-j2\pi(m-1)/M}$ corresponding to the $m$th antenna-element at the receive UCA so that the vortex waves can be transformed into plane waves in digital domain. Then, the sum of recovered azimuthal phase of the $m$th antenna-element at the receive UCA corresponding to OAM-mode ($l_r$) is derived as follows:

$$\sum_{m=1}^{M} \exp\left[j\frac{2\pi(m-1)}{M}l\right]\exp\left[-j\frac{2\pi(m-1)}{M}l_r\right]$$
$$= \sum_{m=1}^{M} \exp\left[j\frac{2\pi(m-1)}{M}(l-l_r)\right]$$
$$= \exp\left[j\frac{2\pi(M-1)}{M}(l-l_r)\right]\frac{\sin 2\pi(l-l_r)}{\sin\frac{2\pi(l-l_r)}{M}}$$
$$= \begin{cases} M & l = \frac{\alpha M}{2}+l_r; \\ 0 & l \neq \frac{\alpha M}{2}+l_r, \end{cases} \quad (25)$$

where $\alpha \in \mathbb{Z}$. We can obtain the same value for different OAM-mode using Eq. (25). Thus, it is impossible to distinguish different OAM-modes and recover the signals with different OAM-modes when $l < (1-M)/2+1$ and $l > M/2$.

In summary, the available OAM-modes for OAM based wireless communications system are given as follows:

$$\min\left\{\left[\frac{2-N}{2}\right],\left[\frac{2-M}{2}\right]\right\} \leq l \leq \min\left\{\left[\frac{N}{2}\right],\left[\frac{M}{2}\right]\right\}. \quad (26)$$

where $N$ and $M$ can be odd numbers in Eq. (26). Then, the maximum DoF for UCA based OAM system is $\min\{N,M\}$. ∎

The maximum DoF, denoted by $\text{rank}(\boldsymbol{H}_{\text{MIMO}})$, for the correlated MIMO based wireless communications is the rank of $\boldsymbol{H}_{\text{MIMO}}$, which is smaller than $\min\{N,M\}$. When the antennas are 100% correlated, the DoF of the correlated MIMO is one. When the correlation relaxes until there is uncorrelation, the DoF eventually increases until $\min\{N,M\}$. Thus, the DoF for the correlated MIMO system is smaller than $\min\{N,M\}$. The exact DoF depends on the correlation.

Then, we can obtain that the ratio, denoted by $\mathcal{R}_{DoF}$, of the DoF related to OAM based wireless communications to the DoF related to MIMO based wireless communications as follows:

$$\mathcal{R}_{DoF} = \frac{\min\{N,M\}}{\text{rank}(\boldsymbol{H}_{\text{MIMO}})}. \quad (27)$$

### C. Capacities for OAM and MIMO Based Wireless Communications

In this subsection, the capacities of OAM based wireless communications and MIMO based wireless communications are presented under the condition that the transmit signals have the same power.

*Case A:* Correlated MIMO Based Wireless Communications

For correlated MIMO based wireless communications, the capacity, denoted by $C_{\text{MIMO}}$, can be derived as follows:

$$C_{\text{MIMO}} = B \log_2\left[\det\left(\boldsymbol{I}+\frac{P}{\sigma_i^2 N}\boldsymbol{H}_{\text{MIMO}}\boldsymbol{H}_{\text{MIMO}}^H\right)\right]$$
$$= \sum_{i=1}^{\text{rank}(\boldsymbol{H}_{\text{MIMO}})} B \log_2\left(1+\frac{P\gamma_i}{\sigma_i^2 N}\right), \quad (28)$$

where $(\cdot)^H$ represents the conjugate transpose operation. The notation $\gamma_i$ represents the singular value for the $i$th subchannel of MIMO system and we can obtain the singular values of a matrix by singular value decomposition. The parameter $\sigma_i^2$ is defined as the variance of received noise corresponding to the $i$th subchannel. We denote by $P$ and $B$ the total power for the MIMO system and the system bandwidth, respectively.

*Case B:* OAM Based Wireless Communications

For $N$ subchannels in OAM based wireless communications, without taking into account the mutual in-terference among different OAM-modes, we can derive the capacity, denoted by $C_{\text{OAM}}$, in Eq. (29), where $\sigma_l^2$ represents the variance of received noise corresponding to the $l$th OAM-mode.

## V. SIMULATION ANALYSIS

In this section, we numerically analyze OAM based wireless communications. First, we show the spatial correlation among the different antenna-elements. Then, we evaluate the ratio of DoF related to MIMO based wireless communications to the DoF related to OAM based wireless communications. Finally, we compare the capacities of MIMO and OAM based wireless communications, where we set the bandwidth as 10 MHz.

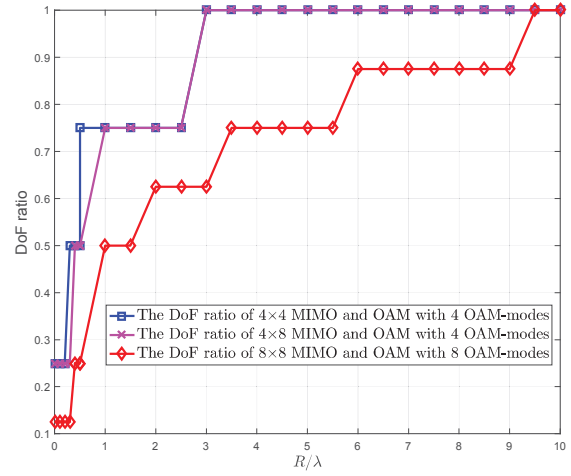

Fig. 3. The DoF ratio of MIMO based wireless communications to OAM based wireless communications.

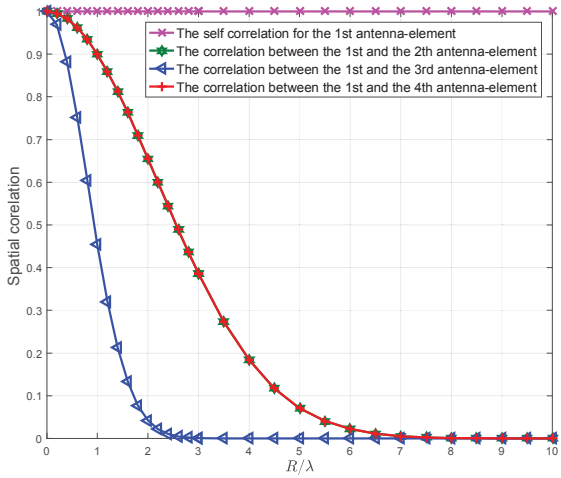

Fig. 2. The spatial correlation among different antenna-elements.

Figure 2 shows the spatial correlation between the 1st antenna-element and all antenna-elements, where four antenna-elements are equipped in the receive UCA. We set the value of $\delta$ and $\vartheta$ as 0.2 and $\pi/6$, respectively. The spatial correlation among different antenna-elements decreases as the value of $R/\lambda$ increases. As we can see from Fig. 2, it is different for the curve corresponding to the spatial correlation of the 1st antenna-element and the 2nd antenna-element and the curve corresponding to the spatial correlation of the 1st antenna-element and the 4th antenna-element. When the value of $R/\lambda$ is not less than 7, we can see that the spatial correlation is zero except the auto-correlation for the 1st antenna-element. In this case, it is the uncorrelated MIMO.

Figure 3 displays the DoF ratio of MIMO based wireless communications to OAM based wireless communications, where we set the the value of $\delta$ and $\vartheta$ as 0.005 and $\pi/6$, respectively. There are three cases in Fig. 3, i.e., the MIMO system with four antenna-elements equipped in both transmitter and receiver and its corresponding OAM system with four OAM-modes, the MIMO system with four antenna-elements equipped in the transmitter and eight antenna-elements equipped in the receiver and its corresponding OAM system with four OAM-modes, and the MIMO system with eight array-elements equipped in both transmitter and receiver and its corresponding OAM system with eight OAM-modes. Because of the orthogonality among different OAM-modes, the DoF of OAM based wireless communications is equal to $\min\{N, M\}$. When the value of $R/\lambda$ is relatively small, the DoF ratio is very small. This is because the DoF of the MIMO based wireless communications is very small. The DoF ratio increases as the value of $R/\lambda$ increases. For the case of the MIMO system with four antenna-elements equipped in both transmitter and receiver and its corresponding OAM system with four OAM-modes, the DoF ratio is equal to 1 when the value of $R/\lambda$ is larger than 3. This is because the increase of $R/\lambda$ relaxes the spatial correlation among different antenna-elements. When the value of $R/\lambda$ is larger than 3, the MIMO system is the uncorrelated MIMO system.

Figure 4 depicts the capacities for the MIMO based wireless communications and the OAM based wireless communications, where we set the value of $R/\lambda$, $\delta$, and $\vartheta$ as 2, 0.2, and $\pi/6$, respectively. In this case, the MIMO system is the correlated MIMO system. We can observe that the capacity of $4 \times 4$ correlated MIMO system is smaller than that of $8 \times 8$ correlated MIMO system. The reason is that the number of subchannels of the $8 \times 8$ correlated MIMO system is greater than that of the $4 \times 4$ correlated MIMO system. For the correlated MIMO system, the capacity linearly increases as the channel SNR increases. For the OAM system, we can see that the capacity of OAM with eight OAM-modes is larger than that of OAM with four OAM-modes. This is because the number of subchannels and capacity of the OAM system increase as the number of OAM-modes increases. For the correlated MIMO system and its corresponding OAM system, we can observe that when the channel SNR is relatively small, the capacity of the correlated MIMO is larger than that of the OAM system. When the channel SNR is relatively large, the capacity of the correlated MIMO is smaller than that of the OAM system. This is because the channel gain for the correlated MIMO system is larger than the channel gain of the OAM system while the number of subchannels for the correlated MIMO is smaller than the number of subchannels for the OAM system. Thus, it is more efficient to use OAM beams to achieve high data transmission when the channel SNR is relatively large.

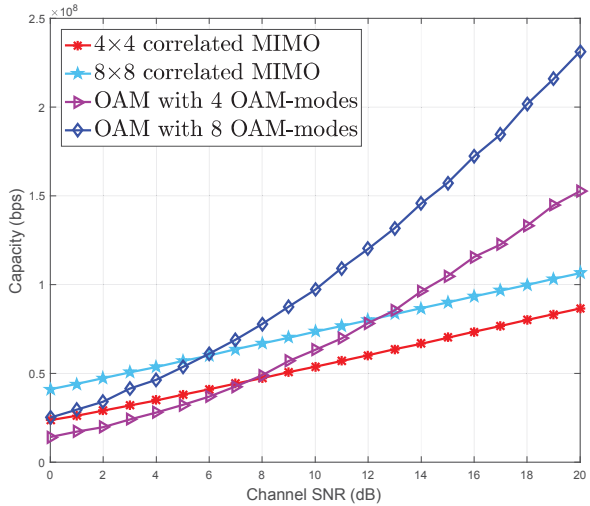

Fig. 4. Capacities for the MIMO and OAM based wireless communications.

## VI. Conclusions

We studied the OAM wireless communications system in terms of orthogonality, DoF, and capacity, where both the transmitter and the receiver are UCA antennas and they are aligned with each other. We compared the DoF and capacity of the OAM based wireless communications with those of the MIMO based wireless communications. The obtained numerical results show that the DoF of the OAM based wireless communications is larger than or equal to the MIMO based wireless communications. When the channel SNR is relatively large, it is better to employ the OAM-modes for high capacity.